\documentclass{article}
\input amssym.def
\input amssym.tex
\usepackage[a4paper]{geometry}
\usepackage{graphicx}
\def\ben{\begin{equation}}
\def\een{\end{equation}}
\def\half{\frac{1}{2}}
\def\bea{\begin{eqnarray}}
\def\eea{\end{eqnarray}}

\def\bx{{\bf x}}

\def\bn{{\bf n}}

\def\p{\partial}
\def\mathbb{\Bbb}

\def\ben{\begin{equation}}
\def\een{\end{equation}}
\def\half{{1 \over 2}}
\def\bea{\begin{eqnarray}}
\def\eea{\end{eqnarray}}
\def\bi{{\bf i}}
\def\bj{{\bf j}}
\def\bn{{\bf n}}
\def \p{\partial}
\def\p{\partial}
\def\babla{{\mbox{\boldmath $\nabla$}}}

\def\x{{\bf x}}
\def\BZ{{\Bbb  Z}}
\def\n{{\bf n}}\def\t{{\bf t}}\def\k{{\bf k}}

\newcommand{\no}{n_o}
\newcommand{\nee}{n_e}

\newcount\hour \newcount\minute
\hour=\time  \divide \hour by 60
\minute=\time
\loop \ifnum \minute > 59 \advance \minute by -60 \repeat
\def\nowtwelve{\ifnum \hour<13 \number\hour:
                      \ifnum \minute<10 0\fi
                      \number\minute
                      \ifnum \hour<12 \ A.M.\else \ P.M.\fi
         \else \advance \hour by -12 \number\hour:
                      \ifnum \minute<10 0\fi
                      \number\minute \ P.M.\fi}
\def\nowtwentyfour{\ifnum \hour<10 0\fi
                \number\hour:
                \ifnum \minute<10 0\fi
                \number\minute}

\title{The helical phase  of chiral nematic liquid crystals as the
  Bianchi $VII_0$ group manifold}
\author{G. W. Gibbons$^1$ and  C.M. Warnick$^{1, 2}$ 
\\
\\ \small{1. D.A.M.T.P., Cambridge, Wilberforce Road, Cambridge CB3 0WA,
  U.K.}
\\ \small{2. Queens' College, Cambridge, CB3 9ET, U.K.} \\}

\begin{document}

\maketitle {\let\thefootnote\relax\footnotetext{{\em Emails}:
  g.w.gibbons@damtp.cam.ac.uk, c.m.warnick@damtp.cam.ac.uk \\ \mbox{} \hspace{.45cm}\emph{Pre-print no.} DAMTP-2010-43}}

\begin{abstract}
We show that the optical structure of the helical phase of a chiral
nematic is naturally associated with the Bianchi $VII_0$ group
manifold, of which we give a full account. The Joets-Ribotta metric governing propagation of the
extraordinary rays is invariant under the simply transitive action of the universal cover $\tilde E(2)$ 
of the three dimensional Euclidean group of two dimensions.
Thus extraordinary light rays are geodesics of a left-invariant
metric on this Bianchi type $VII_0$ group. We are able to solve by separation
of variables both the wave equation and the Hamilton-Jacobi equation
for this metric. The former reduces to Mathieu's equation and the later to the quadrantal
pendulum equation. We discuss Maxwell's equations for uniaxial
optical materials where the configuration is invariant under a group
action and develop a formalism to take advantage of these symmetries.
The material is not assumed to be impedance matched, thus
going beyond the usual scope of transformation optics. We show that 
for a chiral nematic in its helical phase Maxwell's equations
reduce to a generalised Mathieu equation. Our results may also be
relevant to helical phases of some magnetic materials and to light
propagation in certain cosmological models.

\end{abstract}

\section{Introduction}

Recent years have seen a growing interest in
the application of the geometrical ideas originally developed 
for studying  Einstein's theory of 
General Relativity to other
areas of physics, such as condensed matter physics.   
The motivation is both the  theoretical aim
of developing the mathematical tools capable
of dealing with as wide a range of physical problems
as possible, and the  desire to construct laboratory 
analogues of the exotic conditions which general relativity
allows, but which are likely ever to remain inaccessible
to direct experimental investigation. This in turn may provide
a stimulus for further laboratory investigations, possibly resulting
in the discovery of new physical effects.

In the present article we shall pursue this direction by demonstrating
how the mathematical formalism of Lie groups, which is of widespread
use in General Relativity and High Energy Physics \cite{Nakahara, Frankel}, can be harnessed to study
optical properties of symmetrical phases of matter. We develop a
formalism allowing the symmetries of an electromagnetic medium to be
directly exploited in solving Maxwell's equations. When considering symmetries in classical
mechanics, the discussion is simplified by passing to the Lagrangian
or Hamiltonian picture. We present a formalism which similarly
makes symmetries manifest for Maxwell's equations.  In order to
motivate and illuminate the development of this formalism, we consider the example of
light propagation in chiral nematic liquid crystals. We believe this to be the first
application of Lie group techniques to such a problem. The tools we
develop, however, are more widely applicable to media with a
continuous symmetry group. They may also be considered a
generalisation of `transformation optics', extending those ideas to
allow for the possibility that the dielectric tensor and magnetic
susceptibility differ.

We begin in section \ref{sec2} with a brief discussion of the helical
ground state for a chiral nematic liquid crystal. In section
\ref{sec3} we introduce the geometry of the optical metric
of Joets and Ribotta \cite{Joets}, describing
the propagation of the extraordinary ray in a uniaxial birefringent material.
In particular we study this metric and its geodesics
for the helical ground state of a chiral nematic. The resulting
metric is invariant under the simply transitive action
of  three-dimensional
group of isometries which is locally isometric  to
the Euclidean group $E(2)$ of the plane, which we discuss in detail in
section \ref{sec31}.
In fact the  isometry is the universal  cover $\tilde E(2)$ 
and the Lie algebra is of type $VII_0$ in Bianchi's
classification.  We
extend the discussion to $VII_h$ in an appendix. The identification of the symmetry group for the
chiral phase permits a fully geometrical discussion of electromagnetic
phenomena, an approach we exploit.  
The high degree of symmetry allows us to solve the Hamilton-Jacobi
and wave equations up to one quadrature in the former and up 
to solutions of Mathieu's equation  in the latter case.
This  opens up the possibility of 
a detailed analytic investigation of the type of caustics
and optical singularities 
which should be observable in such systems
 \cite{Joets1, Joets2, Joets3,Joets4}. We then go beyond the geometric optics approximation to consider the
full Maxwell equations in Section \ref{sec4}. For a uniaxial
material whose director field takes on a helical configuration we
show how the theory of Lie groups leads to separation of variables for
these equations. The resulting equations take the form of a
generalised Mathieu equation. This result is similar to others in the
literature \cite{peterson, Dreher}, but the derivation is fully motivated from the inherent
symmetries of the problem.

Throughout the paper, we shall make use of the machinery of
differential geometry, in particular tangent vectors, differential
forms and the Lie derivative. References \cite{Nakahara, Frankel}
provide readable accounts of these concepts.

\section{Chiral Nematics and their Helical Ground State \label{sec2}}

A nematic liquid crystal has an order parameter given by
a {\it director} or direction field specified by a unit vector
$\bn =(n_1,n_2,n_3)$, defined up to a sign  $\bn \sim -\bn$, with $\bn
\cdot \bn = 1$.
A {\it chiral}  nematic has a built in twist, specified by a parameter
$q$, which can be realised as a \emph{torsion}, which alters the usual derivative operator acting on a vector by 
\ben
\nabla ^q_i n_j= \nabla _i n_j + q \epsilon_{ijk}n_k\,.
\een
The Frank-Oseen free energy functional in the one-constant
approximation is equivalent, up to a boundary term, to:
\ben
F[\bn] = \half \int \bigl(   | \babla ^q  \bn |^2 - 
\lambda (\bn \cdot \bn -1) \bigr ) d^3\, x   \,,
\een  
where we have  added a  Lagrange multiplier field $\lambda$ to enforce
the constraint that $\bn\cdot \bn=1$. The free energy would be minimised if
\ben
\nabla ^q_i n_j= 0. \label{bog}
\een
However, as can be seen by taking another $\nabla^q$ derivative and 
skew symmetrising,  this is not possible over an extended region so the system is {\it frustrated}  
and must adopt some compromise configuration \cite{Sethna,Sethna1,
Sethna2}.

One such configuration is the helical phase for which
\ben
\bn= \bi \cos (pz) + \bj \sin (pz)\,. 
\een
For more details about the liquid crystals the reader may consult
\cite{Bragg,Collings,Wright,Mermin,Kleman,Chandra}. For the helical phase,
\ben
\babla \cdot \bn=0\,,\qquad \babla \times \bn =-p \bn \,,\qquad \nabla
^2 \bn =-p^2 \bn \,, 
\een
This configuration is a stationary point of the free energy,
satisfying the second order Euler-Lagrange equation resulting from
extremising $F[\bn]$:
\ben
-\nabla ^2 \bn +2 q \nabla \times \bn = (\lambda-2q^2) \bn\,.
\een 
provided we choose
\ben
\lambda=p^2-2 pq+2q^2.
\een
Among these solutions of the Euler-Lagrange equations, the one
minimising the free energy density has $p=q, \lambda=q^2$. The only non-vanishing components of $\nabla ^q_i n_j$ are 
\ben
\nabla^q _2n_3=qn_1\,\qquad \nabla ^q_1  n_3=-q n_2.
\een
It follows that  the helical ground state is not  a solution of the 
first order frustrated  ``Bogomolnyi equation'' 
\ben
\nabla ^q_i n_j= 0.
\een 

Another means of relieving the frustration is the `double twist'
structure, given in cylindrical polar coordinates $(\rho, z, \phi)$ by
\begin{equation}
\bn = \mathbf{e}_{z} \cos q \rho - \mathbf{e}_\phi \sin q \rho.
\end{equation}
Along the $z$ axis, this configuration has $\nabla ^q_i n_j= 0$, so
inside a sufficiently small cylinder, the free energy density is in
fact \emph{lower} than that for the helical phase. A structure composed of these
tubes can fill space, but there will necessarily be defects where the
tubes meet \cite{Sethna2}. Such a configuration gives the so called
`blue phase'. Whether the blue or the helical phase is
thermodynamically preferred depends on the energetic cost associated
to accommodating the defects of the blue phase.

\section{Optical metrics for Nematics \label{sec3}}

If $\n$ is the director, and $\t={d\x \over ds}$,
with $ds^2 = d \x ^2 $ is  the unit tangent vector,
then  then the inverse speed or slowness of an extraordinary ray is 
is given by \cite{Joets}
\ben
n=\sqrt{\no^2 ( \t .\n)^2 + \nee ^2 \bigr (\t - \n (\n \cdot \t)   \bigr )  ^2} \,, 
\een
  where $\no$ is the refractive index of the {\it ordinary ray}  and 
$\nee $ that of the
{\it extra-ordinary ray}.
Fermat's principle reads
\ben
\delta \int n ds =0 \,.
\een
Thus the rays are geodesics of the {\it Joets-Ribotta metric}
\ben
ds _o^2 = \nee ^2 d \x ^2 + (\no^2 -\nee ^2 ) (\n .d \x ) ^2 \,.
\een

Assuming the refractive indices are constants, we can write down the
metric for the helical ground state given above
\begin{eqnarray}
ds _o^2 &=& (\no^2 \cos^2 (pz) + \nee ^2 \sin^2 (pz))dx^2+ \bigl (\no^2 \sin^2
(pz) +\nee ^2 \cos^2 (pz) \bigr )dy^2 \nonumber \\ && + 
(\no^2-\nee ^2)\sin (2 p z)  dx dy + \nee ^2 dz^2 \label{metric}
\end{eqnarray}

As another example, consider a particular case where the director
field, $\bn$, is in a `hedgehog' configuration 
(cf. \cite{Satiro,Pereira,Pereira2}) and where in addition the
refractive indices $\no, \nee$, vary with position inside the ball $r = |\x|<1$ according to
\ben
\n = {\x \over r}\,,\qquad \nee  = {1 \over \sqrt{1-r^2} } \,,\qquad
\no=  {1 \over {1-r^2} } \,. \label{hyp}
\een
The resulting Joets-Ribotta metric is  
\ben
ds_o^2 = {d \x ^2 \over 1-r^2} + {(\x .d \x )^2 \over (1- r^2) ^2 } \,.  
\een
which one recognises as that of Hyperbolic Space $H^3$ 
in Beltrami coordinates. Remarkably, because Hyperbolic space is
projectively flat in these coordinates,  the light rays are straight lines
in this case. One could consider the 3-sphere $S^3$, by changing the
sign in front of $r^2$. Then one has a model related to Maxwell's
fish-eye lens by a coordinate transformation, but whose rays are
straight lines. Other examples may be found in
\cite{Satiro,Pereira,Pereira2}. Optical properties such as (\ref{hyp})
my appear unnatural, but modern meta-materials are increasingly able
to mimic such refractive indices, at least within a certain range of
frequencies for the electromagnetic field.
  
\subsection{$E(3)$ and Left-invariant metrics \label{sec31}}

The aim of the present section is to obtain
the isometry group of the apparently quite complicated  metric (\ref{metric}).
An enormous  simplification results if we use the formalism
of metrics on Lie groups. We will start with a brief discussion,
tailored to the Euclidean group $E(2)$ of isometries of the plane,
consisting of translations and rotations in two dimensions. Those
familiar with the construction of left- and right-invariant forms on Lie
groups may wish to skip to the summary at the end of this subsection.

We can realise elements of $E(2)$ as matrices as follows. We first fix
a number $p$. To any
point $(X,Y)$ in the plane, we associate the column vector
\begin{equation}
(X,Y) \sim \left(\begin{array}{c}X \\ Y \\ 1 \end{array} \right).
\end{equation}
We first note that a general isometry of the plane can be decomposed into a clockwise rotation
of angle $p z$
about the origin, followed by a translation of $(x,y)$. We can write
down a matrix $M(x,y,z)$ depending on parameters $(x,y,z)$ which performs this operation as follows
\begin{equation}
\left(\begin{array}{c}X' \\ Y' \\ 1 \end{array} \right) =
\left(\begin{array}{ccc}\cos pz & -\sin pz & x \\ \sin pz & \cos pz &
    y \\ 0&0&1 \end{array} \right) \left(\begin{array}{c}X \\ Y \\ 1
  \end{array} \right) = M(x,y,z) \left(\begin{array}{c}X \\ Y \\ 1
  \end{array} \right). 
\end{equation}
Corresponding to any three numbers $(x,y,z)$, we have a unique
isometry and conversely each isometry corresponds to a unique
$(x,y,z)$, provided that we regard $z$ and $z+2\pi/p$ as the
same\footnote{for most of the rest of the paper, we shall in fact work
  with the covering group $\tilde{E}(2)$ obtained by dropping the
  identification of the $z$ coordinate. This is a slightly technical
  point which we shall not labour.}. We
refer to $(x,y,z)$ as coordinates on the group $E(2)$. 

Our aim is to construct a set of vector fields which are invariant
under an action of the group $E(2)$. We'll consider for the moment a
matrix Lie group $\mathcal{G}$, i.e.\ a group whose elements are $n\times n$
matrices for some appropriate $n$ where the group action is by matrix
multiplication\footnote{There are some further assumptions on the
  smoothness of the group, but we'll take these as given.}. We note that any element $A$ of
 $\mathcal{G}$ gives rise to two natural transformations on the group
itself. Acting on $M$, a general element of $\mathcal{G}$, they give:
\begin{equation}
\Lambda_{A}(M) = A M, \qquad P_{A}(M) = MA
\end{equation}
and are known respectively as left and right translation, or the left
and right action of $A$. We note that the left action and the right action commute:
\begin{equation}
\Lambda_{A}(P_B(M))=\Lambda_{A}(MB) = A M B = (\Lambda_{A}(M)) B = P_B(\Lambda_{A}(M)). \label{comact}
\end{equation}
Now suppose that $A$ is infinitesimally close to the identity
matrix, $A = I + \epsilon \delta A$ for an infinitesimal parameter $\epsilon$. We have
\begin{equation}
\Lambda_{A}(M) = (I+\epsilon \delta A) M = M + \epsilon \delta_L M
\end{equation}
where the infinitesimal generator $\delta_\Lambda M = (\delta A) M$ can be interpreted as a tangent vector
to $\mathcal{G}$ at the point $M$, where we think of $\mathcal{G}$ as
a submanifold, i.e.\ a surface\footnote{We do not assume that a
  surface is necessarily $2$-dimensions, merely that it is of lower
  dimension than the space in which it lives.} in the space of all $n \times n$ matrices. This is a \emph{right-invariant}
vector field, since $\delta_\Lambda (P_B(M)) = P_B(\delta_\Lambda M)$ because of equation
(\ref{comact}) above. Following a similar procedure, we find that the
vector fields $\delta_P M = M (\delta A)$ are \emph{left-invariant}.

In order to construct the left- and right-invariant vector fields of
$E(2)$, we must first find all suitable $\delta A$ such that $I+\epsilon \delta A$
is an element of $E(2)$. We can do this by using the coordinate
representation $M(x,y,z)$ above. Since $M(0,0,0)=I$, we can find the
most general $\delta A$ by Taylor expanding $M(x,y,z)$ for small
$(x,y,z)$. We find that the general $\delta A$ is a linear combination
of the matrices:
\begin{equation}
M_1 = \left(\begin{array}{ccc}
0&0&1\\0&0&0\\0&0&0
\end{array}\right) \qquad M_2 = \left(\begin{array}{ccc}
0&0&0\\0&0&1\\0&0&0
\end{array}\right)\qquad M_3 = \left(\begin{array}{ccc}
0&-1&0\\1&0&0\\0&0&0
\end{array}\right).
\end{equation}
We'd like to express the vector fields in terms of the $(x,y,z)$
coordinates. To do this, we define a basis of vector fields for $E(2)$
as follows
\begin{equation}
\frac{\partial}{\partial x} = \frac{\partial M(x,y,z)}{\partial x},
\quad \frac{\partial}{\partial y} = \frac{\partial M(x,y,z)}{\partial y}, \quad \frac{\partial}{\partial z} =  \frac{\partial M(x,y,z)}{\partial z}.
\end{equation}
We can either think of $\partial / \partial x$ as a concrete matrix
tangent to $E(2)$ as a surface in the space of $3 \times 3$ matrices,
or more abstractly as the vector field which generates a shift from
$(x,y,z)$ to $(x+\delta x, y, z)$ in the coordinate space. This
notation captures the fact that under a change of variables for the
coordinate space, $(x, y, z) \to (x', y', z')$, the vector fields
transform in the same way as differential operators following the
chain rule. We can readily
calculate the left invariant vector field corresponding to $M_1$:
\begin{equation}
L_1 = M M_1 = \left(\begin{array}{ccc}
0&0&\cos p z\\0&0&\sin p z\\0&0&0
\end{array}\right) = \cos p z \frac{\partial}{\partial x} + \sin p z  \frac{\partial}{\partial y}.
\end{equation}
In a similar way, we can find the rest of the left- and
right-invariant vector fields, $L_i = M M_i, R_i = M_i M$:
\begin{equation}
\begin{array}{rclcrcl}
L_1 &=&  \cos p z \frac{\partial}{\partial x} + \sin p z
\frac{\partial}{\partial y}, & \qquad & R_1&=& \frac{\p}{\p x},  \\
L_2 &=& \cos pz \frac{\p}{\p y}-  \sin pz 
\frac{\p}{\p x},&\qquad & R_2&=& \frac{\p}{\p y}, \\
L_3 &=& \frac{1}{p} \frac{\p}{\p z}, &\qquad & R_3 &=& \frac{1}{p}\frac{\p}{\p z} + x \frac{\p}{\p y} -y  \frac{\p}{\p x}.
\end{array}
\end{equation}
Now that we have the left- and right-invariant vector fields, we can
construct the left- and right-invariant one-forms which are dual to
them. Taking $dx, dy, dz$ to be the one-forms dual to $\p /\p x,
\p/\p y, \p/\p z$, the left-invariant forms $\lambda^i$ and right
invariant forms $\rho^i$ are:
\begin{equation}
\begin{array}{rclcrcl}
\lambda^1 &=& \cos (pz) dx + \sin (pz) dy, & \qquad & \rho^1&=& dx +
py dz, \\ \lambda ^2 &=& \cos (pz) dy   - \sin (pz) dx, &\qquad&
\rho^2&=& dy - px dz \\ \lambda ^3&=& p dz, &\qquad &\rho^3&=& p dz.
\end{array}
\end{equation}

The matrices $M_i$ have a natural commutator algebra, the Lie algebra
$\frak{e}(2)$. This determines in a natural way the Lie algebra of the
vector fields $L_i$ and $R_i$:
\begin{equation}
\begin{array}{rclcrcl}
\bigl [ L_1 , L_2 \bigr ] &=& 0, &\qquad& \bigl [ R_1 , R_2 \bigr ] &=& 0,\\
\bigl [ L_3 , L_1 \bigr ] &=& +L_2,  &\qquad&\bigl [ R_3 , R_1 \bigr ] &=& -R_2, \\
\bigl [ L_3 , L_2 \bigr ]  &=& -L_1,  &\qquad&\bigl [ R_3 , R_2 \bigr ]  &=& +R_1,
\end{array}
\end{equation}
and the Maurer-Cartan algebra of the one-forms
\begin{equation}
\begin{array}{rclcrcl}
d \lambda ^1&=&+ \lambda ^3 \wedge \lambda ^2\,,&\qquad&d \rho ^1&=&- \rho ^3 \wedge \rho ^2\,,\\
d \lambda ^2&=& -\lambda ^3 \wedge \lambda ^1 \,,&\qquad&d \rho ^2&=&
+\rho ^3 \wedge \rho ^1 \,, \\
d \lambda ^3&=& 0\,,&\qquad&d \rho ^3&=& 0\,.
\end{array}
\end{equation}
The fact that left and right actions commute (\ref{comact}) is
reflected in the fact that $[L_i, R_j]=0$.

We can check the claimed invariance explicitly. First we note the
matrix identity
\begin{equation}
M(x,y,z) M(\nu, \eta, \zeta) = M(x+\nu \cos p z -\eta \sin p z, y+\eta
\cos p z + \nu \sin p z, z+ \zeta), \label{lraction}
\end{equation}
from which we deduce that the element $M(\nu, \eta, \zeta)$ acting by
right translation takes $(x,y,z)$ to $(x', y', z')$ where
\bea
x'&=& x+\nu \cos p z -\eta \sin p z, \\ y'&=& y+\eta
\cos p z + \nu \sin p z,\\ z' &=& z+ \zeta.
\eea
Making these substitutions into $\rho^i$, treating $\nu, \eta, \zeta$
as constants, we find
\bea
dx +py dz&=&dx' +py' dz', \\
dy - px dz &=& dy' - px' dz', \\
p dz &=& p dz',
\eea
so that the $\rho^i$ are indeed invariant under right
translations. Interchanging the roles of $x,y,z$ and $\nu, \eta,
\zeta$ in (\ref{lraction}) we deduce that the element $M(\nu, \eta, \zeta)$ acting by
left translation takes $(x,y,z)$ to $(x', y', z')$ where
\bea
x'&=& x \cos p \zeta -y \sin p \zeta+\nu, \\ y'&=& y\cos p \zeta +x \sin p \zeta+\eta,\\ z' &=& z+ \zeta.
\eea
substituting into $\lambda^i$, again treating $\nu, \eta, \zeta$
as constants, we can check that
\bea
\cos (pz) dx + \sin (pz) dy&=&\cos (pz') dx' + \sin (pz') dy', \\
\cos (pz) dy - \sin (pz) dx&=&\cos (pz') dy' - \sin (pz') dx', \\
p dz &=& p dz',
\eea
so the $\lambda^i$ are invariant under left translations.

Armed with these invariant one-forms, we are now in a position to
construct metrics which are invariant under an action of $E(2)$. For
example, the
flat metric can be written in terms of the left-invariant one-forms as: 
\ben
ds ^2 = p^{-2} (\lambda ^3 )^2 + (\lambda ^1)^2 + (\lambda ^2) ^2 ,
\een
and is hence manifestly left-invariant. Now for the helical ground state of the nematic liquid crystal,
\ben
\bn \cdot d \x = \lambda ^1\,,
\een
thus the {\it Joets-Ribotta metric} of the helical phase may be written as 
\bea
ds ^2_o &=&  \nee ^2 \bigl(  p^{-2} (\lambda ^3 )^2 + (\lambda ^1)^2 +
(\lambda ^2) ^2\bigr ) + (\no^2-\nee ^2)( \lambda ^1 )^2 \,, \nonumber
\\
&=& \no^2 (\lambda^1)^2 + \nee ^2 (\lambda^2)^2 +
\frac{\nee ^2}{p^2}(\lambda^3)^2 \label{JR}
\eea
which is a left invariant metric on $\tilde E(2)$. In fact, any
left invariant metric may be brought into this form by a global right
action of $E(2)$.
Cartan's  formula
for a p-form  reads
\ben
{\cal L}_X \, \omega = i_X d \omega + d(i_X \omega) \,,  
\een
and so the non-vanishing Lie derivatives are 
\bea
{\cal L }_{L_3} \, \lambda ^1      &=&- \lambda ^2 \,,\\ 
{\cal L} _{L_3} \, \lambda ^2 &=& \lambda ^1 \,,\\ 
{\cal L} _{L_1} \, \lambda ^2 &=& -\lambda ^3 \,,\\ 
{\cal L} _{L_2} \, \lambda ^1 &=& \lambda ^3 \,. 
\eea
Thus while $L_3$ is an additional symmetry of the flat metric
none of the $L_i$ are symmetries of the Joets-Ribotta metric.

To summarise then,   $(x,y,z) \in {\Bbb R} ^3 $  may be considered as coordinates  on
$\tilde E(2)$, the universal cover of the two-dimensional
Euclidean group $E(2)$ with  $\lambda ^i$ left-invariant one-forms
and $L_i$ left-invariant vector  fields. If we were to identify
the coordinate modulo  $\frac{2\pi n }{p}$, $ n=1,2,\dots$ 
the group would be the $n$ -fold cover of the Euclidean group
$E(2)$, which corresponds to $n=1$. With this identification, the
Joets-Ribotta metric of the helical phase is a left-invariant metric.
Its symmetry algebra, the Lie algebra $\frak{e} (2)$, is  of Bianchi type $VII_0$.  
As an aside, this may be obtained from the rotation group algebra $\frak{so}(3)$
by means of a Wigner-In\"on\"u contraction.
If $\tilde M_i $ are the generators of $\frak{so}(3)$, one
sets $\tilde M_1 = \frac{1}{\epsilon}  M_1\,, \tilde M_2 = 
\frac{1}{\epsilon}  M_2 \,, \tilde M_3 = M_3$ and takes the limit
$\epsilon \rightarrow 0$.
 Under this contraction, the direction field 
$\bn \cdot d \x = L_1$ arises as the image of the Hopf
fibration generated by the right  action of $\tilde L_1$ 
and the Joets-Ribotta metric as the image of the Berger-Sphere.
For the  relevance of the Hopf fibration to chiral nematics
see  \cite{Sethna,Sethna1,Sethna2,Mermin,Bouligand,Pansu1,Pansu2,Dandoloff}.   
  
\subsection{The Ray Approximation} 

In the ray approximation we are looking 
at geodesics with respect to a left-invariant metric
on the universal cover of the Euclidean group $\tilde E(2)$. 
Now the Euclidean group $E(2)$ is the configuration space
for a rigid body  in two-dimensional
Euclidean space ${\Bbb E} ^2$. The motion of a rigid body
moving in a homogeneous, incompressible, inviscid fluid
\cite{Lamb} 
is known to correspond to  geodesic motion with respect
to a  left-invariant metric on the Euclidean group \cite{Birkhoff,Olshanetsky}.
The present situation corresponds to a cylinder
with its axis in a plane
\cite{Lamb,Ramsey} which may be reduced to the quadrantal
pendulum.     
  
To see this in detail note that  
the Eikonal equation is 
\ben
\frac{p^2} {\nee ^2} (L_3 W)^2  
+ \frac{1}{\nee ^2}( L_2 W )^2  + \frac{1} {\no^2} ( L_1 W) =\omega^2\,,
\een
and it separates. If $W= k_x x + k_yy + G(z) $, then
\ben
\frac{1}{\nee ^2} (\frac{dG}{dz}) ^2 + \frac{1}{\no^2} 
(k_x\cos(pz) + k_y \sin (pz)  )^2 + \frac{1}{\nee ^2} ( k_x\sin (pz) -  k_y \cos(pz)) ^2 =\omega^2\,.
\een 
The Killing vectors $R_i$ give rise to  three constants of the motion
of the form
\ben
p_i = g_{\mu \nu} \frac{dx ^\mu }{dt} R^\mu_i \,.
\een
of which two, $p_1 $ and $p_2$ mutually commute. 

\begin{figure}
\centering
\includegraphics[width=3in]{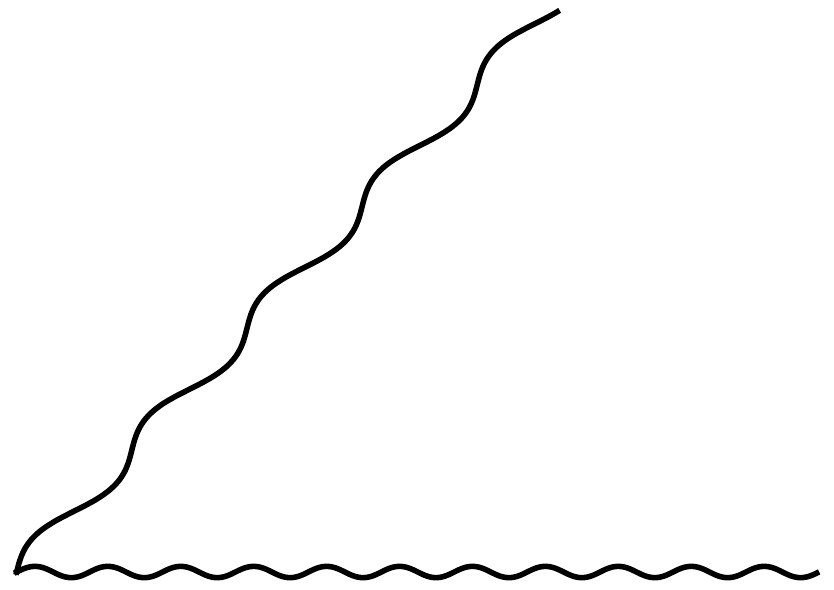}
\caption{Two geodesics of the Joets-Ribotta metric (\ref{metric}). The curves are shown projected into the $x-z$ plane. One is unbounded in $z$, whereas the other is bounded. Not shown is the motion in the $y$ direction which gives both these curves a cork-screw motion. \label{geodfig}}
\end{figure}

We may immediately find the equations for rays in a first order form by making use of the relation:
\begin{equation}
\frac{dx^\mu}{dt} = g^{\mu \nu}\frac{\partial W}{\partial x^\mu}.
\end{equation}
We find:
\begin{eqnarray}
\nee ^2 \dot{x} &=& \frac{k_x}{2}\left(1+\frac{\nee ^2}{\no^2} \right)- \frac{k}{2}\left(1-\frac{\nee ^2}{\no^2} \right) \cos (2 p z - \psi) ,\label{xeq}\\
\nee ^2 \dot{y} &=&  \frac{k_y}{2}\left(1+\frac{\nee ^2}{\no^2} \right)- \frac{k}{2}\left(1-\frac{\nee ^2}{\no^2} \right) \sin (2 p z - \psi) , \label{yeq} \\
\nee ^4 \dot{z}^2&=&  \omega ^2 \nee ^2 
-\frac{k^2}{2} \left(1+ \frac{\nee ^2}{ \no^2} \right) + \frac{k^2}{2}
\left (1- \frac{\nee ^2}{ \no^2} \right ) \cos(2pz-\theta)  \label{zeq}
  .
\end{eqnarray}
Here $\tan \psi = k_y/k_x$, $k = \sqrt{k_x^2+k_y^2}$ and $\tan \theta
= 2k_xk_y/(k_x^2-k_y^2)$. Let us first consider
(\ref{zeq}). Introducing new constants $\alpha, \beta$ and defining $\zeta=pz-\theta/2$, we find
\begin{equation}
\dot{\zeta}^2 -  \frac{p^2}{\nee ^4}(\alpha + \beta \cos(2\zeta)) =0
\end{equation}
which is the so-called {\it quadrantal  pendulum}
 equation \cite{Ramsey}. The pendulum has two different types of
 behaviour, depending on the constants $\alpha$ and $\beta$. If
 $\alpha> |\beta|$, then $|\zeta|$ and hence $|z|$ will increase without
 bound. This corresponds to a pendulum swinging through complete
 revolutions. If $\alpha<|\beta|$, then $\zeta$ will oscillate about $2
 n \pi$ for some integer $n$. This corresponds to the standard
 libratory motion of a pendulum. Thus we find two behaviours for the
 rays. Either the rays can penetrate in the $z$-direction or else they
 are trapped to move between two planes perpendicular to the
 $z$-axis. Finally, we can consider the other equations of motion,
 (\ref{xeq}) and (\ref{yeq}). We can interpret these as saying that
 the tangent vector of the ray oscillates around an average
 direction. For rays which are not bounded in $z$, the result is a
 `cork-screw' curve, similar to a helix. The `tightness' of the spiral
 is determined by how close $\nee ^2/\no^2$ is to $1$. Figure
 \ref{geodfig} shows some examples.

\subsection{The wave equation} 

For an  approximate description of light propagation beyond the ray approximation
one may use the scalar wave equation. The scalar wave equation
captures some of the wave aspects of light, whilst ignoring the
complications relating to polarisation which arise in the full Maxwell
equations. We may expect the wave equation on the Joets-Ribotta metric
to share some features with one of the two polarisations of the full
Maxwell equations. It takes the form:
\ben
0  = -\frac{\p ^2 \Psi}{\p t ^2 } + \frac{p^2} {\nee ^2} L_3 L_3 \Psi 
+ \frac{1}{\nee ^2} L_2 L_2 \Psi + \frac{1} {\no^2} L_1 L_1 \Psi\,. \label{waveeqn}
\een
It separates. That is if $\Psi = e^{i( + k_x x + k_y y -\omega t )} F(z)$, then
\ben
\frac{1}{\nee ^2}  \frac{d^2 F}{dz ^2 } +\Bigl( \omega ^2 - \frac{1}{\no^2}
(k_x\cos(pz) + k_y \sin (pz)  )^2 - \frac{1}{\nee ^2} ( k_x\sin (pz) -  k_y \cos(pz)) ^2 \Bigr ) F =0\,. 
\een 
Recalling $\alpha, \beta, \zeta$ from the previous section, this is of the form
\ben
\frac{d^2 F}{d\zeta ^2 } +\Bigl( \alpha + \beta \cos (2 \zeta) \Bigr ) F=0\,,
\label{Mathieu} \een
which is Mathieu's equation.

By the Floquet-Bloch theorem, the general solution of (\ref{Mathieu}) is of the form
\ben
F= c_1 e^{i \mu \zeta} f(\zeta) + c_2 e^{-i \mu \zeta} f(-\zeta) 
\een
where $f(\zeta)= f(\zeta+2\pi)$ and $\mu$ depends on $\alpha$ and
$\beta$. Expanding $f(\zeta)$ as a Fourier series, we deduce the
Laue-Bragg conditions
that an incoming wave with wave vector $\k_{\rm in}$ incident on some
region where propagation is described by (\ref{waveeqn})
is reflected/diffracted  with wave vector $\k_{\rm out}$
where 
\ben
p (\k_{\rm out}- \k_{\rm in}) _z = m \in {\BZ} \,.
\een
We may think of $\mu=\mu(k_x, k_y, \omega)$ as defining a
dispersion relation, averaged over the period in the vertical
direction. When $\mu$ is real we expect propagating waves, whereas
when $\mu$ has an imaginary component the solutions either decay or
grow exponentially in $\zeta$. It can be shown that the marginal cases
between propagation and damping occur when $\mu=0, \pi$ (note $\mu$ is
only defined up to multiples of $2 \pi$). This defines a set of
surfaces in the $(k_x, k_y, \omega)$ space which separate out the
regions where the wave propagates and where it is damped. To determine
these surfaces, we can (for $\mu=0$, the other case follows similarly)
expand $F$ in  Fourier series:
\ben
F = \sum_{-\infty}^{ \infty}  c_n e^{in\zeta} 
\een
and obtain a three term recurrence relation 
\ben
-n^2 c_n + \alpha c_n  + 2 \beta (c_{n-2}+ c_{n+2} ) =0\,.
\een
The condition that this relation admits a non-trivial solution may be
related to the vanishing of an infinite determinant, a procedure known as
Hill's method, see e.g.\ \cite{WW}.

\section{Maxwell's equations \label{sec4}}

Before we discuss Maxwell's equations for the helical phase of a
nematic liquid crystal, we will first formulate Maxwell's equations
for a general medium in the language of differential forms.  This will
be the most convenient language in which to discuss how to apply the
machinery of Lie groups to the problem in hand. We will work directly
with the fields rather than introducing potentials as this avoids
tackling the issue of gauge invariance. We'll work on a
$4$-dimensional manifold $M$, but this restriction is not necessary. We begin by noting
that we can define two $2$-forms: 
\begin{eqnarray}
F &=& \frac{1}{2}F_{\mu \nu}dx^\mu \wedge dx^\nu = E_i dx^i \wedge dt + \frac{1}{2}\epsilon_{ijk}B^i dx^j \wedge dx^k, \nonumber \\
G &=&\frac{1}{2}G_{\mu \nu}dx^\mu \wedge dx^\nu= H_i dx^i \wedge dt - \frac{1}{2}\epsilon_{ijk}D^i dx^j \wedge dx^k. \nonumber
\end{eqnarray}
These forms encode the fields contained in the antisymmetric
$4$-tensors with components
\begin{equation}
\begin{array}{rclcrcl}
F_{i0} &=& E_i &\quad& F_{ij} &=& \epsilon_{ijk} B^k \\
G_{i0} &=& H_i &\quad& G_{ij} &=& -\epsilon_{ijk} D^k 
\end{array}
\end{equation}
Where $E_i$, $B^i$ are the electric field and magnetic displacement, $D^i$ is the
electric displacement field and $H_i$ is the magnetic field. The
advantage of packaging the fields as two-forms is that Maxwell's
equations become the simple pair of relations
\begin{equation}
dF = 0, \qquad dG = J, \label{max2}
\end{equation}
with $J$ the current $3$-form. Of course, to close this system of
equations for a prescribed $J$, we must specify a relation between $F$
and $G$, the \emph{constitutive relation}. This is nothing more than
the usual relations one requires relating $(\mathbf{E}, \mathbf{B})$
and $(\mathbf{D}, \mathbf{H})$. In the language of forms, we require
a map from the space of sections of $\Omega^2(M)$ to
itself\footnote{In higher dimensions, $G$ will be a $n-2$ form, but
  similar considerations apply}. In many materials, the constitutive
relation is local and linear, so may be represented by a section of
the bundle $End(\Omega^2(M)$, i.e. a possibly space dependent linear map $C$ such that
\begin{equation}
G = CF,\quad \textrm{i.e.}\quad G_{\mu\nu}=C_{\mu \nu}{}^{\kappa
  \tau}F_{\kappa \tau}
\end{equation}
where $C$ acts pointwise. This tensor $C$, together with the
differentiable structure of the manifold, is the minimal data required
to define Maxwell's equations - it has not thus far been necessary to
introduce a metric or other structure to $M$. In index notation the
Maxwell equations take the form:
\ben
\partial_{[\mu} F_{\nu \sigma]} = 0, \qquad
\partial_{[\mu}\left(C_{\nu \sigma]}{}^{\kappa \tau} F_{\kappa \tau} \right)=0.
\een
We note that defining $C$ as an endomorphism, i.e.\ with two indices
up and two down, ensures that it is not necessary to define a
connection in order to take derivatives covariantly.

In order that (\ref{max2}) defines a suitable hyperbolic system of
partial differential equations, restrictions are
required on $C$. We shall assume that $C$ satisfies some such suitable
conditions, without specifying what those might be. As a simple example, we may take $C$ to be the Hodge map
induced by a Lorentzian metric $g$, i.e.\ we take $C = \star_g$. If
$g$ is the flat metric, this gives the classical Maxwell equations in
the vacuum. If $g$ is not flat, we may interpret the field as an
electromagnetic field propagating in a gravitational background. In
the case that $g$ is static, we can alternatively interpret the field
as propagating through some material with a position dependent
dielectric tensor, $\varepsilon_{ij}$ and magnetic susceptibility
$\mu_{ij}$. This is the basis of \emph{transformation optics}
\cite{Tamm, Plebanski, Leonhardt}. For a material in which Maxwell's
equations have a gravitational interpretation it must be the case that
$\varepsilon_{ij}=\mu_{ij}$ in suitable units, i.e.\ the material is
\emph{impedance matched} \cite{Leonhardt}. This need not be the case for a general
material. We will take $C$ to have the following form:
\begin{eqnarray}
C(dx^i \wedge dt) = -\frac{1}{2}\varepsilon_{ij}\epsilon_{jkl} dx^k \wedge dx^l \nonumber \\
C(dx^i \wedge dx^j) = \epsilon_{ijk}(\mu^{-1})_{kl} dx^l \wedge dt. \label{standardC}
\end{eqnarray}
Note that if $C=\star_g$ for some Lorentzian metric, we have $C^2 =
-1$ so that $\varepsilon_{ij}(\mu^{-1})_{jk} = \delta_{jk}$,
justifying our assertion that materials with a gravitational analogue
are impedance matched.

The liquid crystals in which we are interested are \emph{uniaxial}, so
that at each point, we can assume that the quadrics defined by
$\epsilon$ and $\mu$ are \emph{spheroidal} (i.e.\ ellipsoids with an
axis of symmetry) with a common axis. In other
words, there is locally a basis in which the tensors have the form 
\begin{equation}
\varepsilon = \left( {\begin{array}{ccc}
 \varepsilon_{||} & 0&0 \\
 0& \varepsilon_\perp&0 \\
  0&0& \varepsilon_\perp \\
 \end{array} } \right) \qquad \mu = \left( {\begin{array}{ccc}
 \mu_{||} & 0&0 \\
 0& \mu_\perp&0 \\
  0&0& \mu_\perp \\
 \end{array} } \right). \label{biax}
\end{equation}
If we assume that the axis of the material lies along $\bn$, this can be written in a more covariant form as
\begin{equation}
\varepsilon_{ij} = \varepsilon_\perp \delta_{ij} + (\varepsilon_{||}-\varepsilon_\perp)n_in_j, \qquad \mu_{ij} = \mu_\perp \delta_{ij} + (\mu_{||}-\mu_\perp)n_in_j.
\end{equation}
Before we discuss the consequences of such a constitutive relation in the case of a nematic liquid crystal in the helical ground state, let us first consider for a moment the geometric optics approximation.

\subsection{Geometric optics}

Let us consider the Maxwell equations described above in a geometric optics limit. We consider a field which takes the form
\begin{equation}
F = e^{\frac{iS}{\alpha}} \left(F_0 + \alpha F_1 + \ldots \right)
\end{equation}
where by assumption $F_i$ are $O(1)$ as $\alpha \to 0$. We assume that
there are no currents or charges, so that Maxwell's equations become
$dF=dG=0$. We also assume that $C$ varies slowly by comparison to the
wavelength of the field. Inserting our ansatz and collecting terms in
$\alpha$, we find 
\begin{eqnarray}
0&=&\frac{i}{\alpha} dS \wedge F_0 + \sum_{k=1}^\infty (dF_{k-1}+i dS \wedge F_k)\alpha^{k-1} ,\nonumber \\
0&=&\frac{i}{\alpha} dS \wedge CF_0 + \sum_{k=1}^\infty (dCF_{k-1}+i dS \wedge CF_k)\alpha^{k-1}.
\end{eqnarray}
Let us first consider the $O(\alpha^{-1})$ terms. This is a system
which asserts that $F_0$ is in the kernel of a linear operator which
maps from one $6$-dimensional space to another $6$-dimensional
space. The condition that a non-trivial $F_0$ exists gives a
differential condition on $S$ involving $C$ which we interpret as the
eikonal equation. Associated to a solution of the eikonal equation is
a $2$-form $F_0$ which gives the polarisation of the wave. In general,
there will be only one polarisation associated to each solution of the
eikonal equation. Once we have solved for $S$ and $F_0$, we can
inductively construct $F_k$ by solving the equations 
\begin{eqnarray}
0&=&dF_{k-1}+i dS \wedge F_k ,\nonumber \\
0&=&dCF_{k-1}+i dS \wedge CF_k.
\end{eqnarray}
Presumably the well posedness of this system is a necessary condition that $C$ be an acceptable constitutive map.

In the case where $C=\star_g$, the eikonal equation can be shown to reduce to
\begin{equation}
dS \wedge \star_g dS = 0,
\end{equation}
which is the Hamilton-Jacobi equation for geodesics of the metric. In this case, there is a two dimensional space of possible polarisation tensors. They take the form
\begin{equation}
F_0 = dS \wedge f_0, \qquad g(dS, f_0) = 0.
\end{equation}
The Hamilton-Jacobi equation requires that $dS$ be null. Suppose for example that at a point, $dS$ is parallel to $dt-dx$, then the space of polarisations at that point is spanned by $dS \wedge dy$ and $dS \wedge dz$.

In the case where $C$ has the uniaxial form introduced above, the eikonal equation reduces to the form:
\begin{equation}
\left( -\mu_\perp S_t^2 + \frac{1}{\varepsilon_{||}}\nabla S^2 + \left(\frac{1}{\varepsilon_{\perp}}-\frac{1}{\varepsilon_{||}}\right)(\bn \cdot \nabla S)^2 \right) \left( -\varepsilon_\perp S_t^2 + \frac{1}{\mu_{||}}\nabla S^2 + \left(\frac{1}{\mu_{\perp}}-\frac{1}{\mu_{||}}\right)(\bn \cdot \nabla S)^2 \right)= 0
\end{equation}
The medium is thus \emph{birefringent}. We see straight away that the condition on $S$ factors into two separate Hamilton-Jacobi equations associated to the two metrics
\begin{eqnarray}
g_B &=& -\frac{dt^2}{\mu_\perp} + \varepsilon_{||} d\bx^2 + (\varepsilon_\perp-\varepsilon_{||})(\bn \cdot d\bx)^2, \\
g_E &=& -\frac{dt^2}{\varepsilon_\perp} + \mu_{||} d\bx^2 + (\mu_\perp-\mu_{||})(\bn \cdot d\bx)^2 .
\end{eqnarray}
These are both of the Joets-Ribotta form we have previously
considered. It can be checked that the polarisation tensor associated
to a solution of the Hamilton-Jacobi equation of $g_B$ has
$\epsilon_{ijk} F_{ij} n_k=B_\bn=0$, whereas for a solution of the
Hamilton-Jacobi equation of $g_E$, the polarisation tensor has
$n_iF_{it}=E_\bn=0$. Note that we do not require that $\bn$ remains
constant for this derivation, provided it varies slowly compared to
the wavelength of the light. In the case that $\bn$ varies from point
to point, the polarisation will also change so that to leading order
in $\alpha$, either the magnetic or electric field parallel to the
director will vanish, depending on which type of ray we
consider. Often, one takes $\mu_\perp=\mu_{||}$ in which case, $g_E$
is simply the Minkowski metric and its geodesics are the
\emph{ordinary rays}. The rays of the metric $g_B$ are the
\emph{extraordinary rays} and $g_B$ is the Joets-Ribotta metric, where
we identify $\varepsilon_\perp \mu_\perp = \no^2$ and
$\varepsilon_{||} \mu_\perp = \nee^2$. 

If $C$ is of the form (\ref{standardC}), but with no uniaxial
assumption, then the rays will typically be geodesics of a Finsler
geometry. 

\subsection{Symmetries}

So far, we have re-cast familiar results into the notation of
differential forms. Whilst this is a satisfying exercise, it is not
clear that it introduces any benefits beyond putting the
equations in a manifestly coordinate invariant form. For our purposes,
the great advantage is that this form of the equations permits a
concise discussion of the symmetries of the system and allows the
machinery Lie groups to be brought to bear. We start by defining a
Killing vector $K$ to be a vector which satisfies 
\begin{equation}
\mathcal{L}_K C = 0.
\end{equation}
Recall that $C$ is simply a tensor, so the Lie derivative is defined
as a consequence of the differentiable structure of $M$. Making use of
this and Cartan's relation, we deduce that if a $2$-form $F$ obeys
Maxwell's equations:
\begin{equation}
dF = 0, \qquad d(CF) = 0,
\end{equation}
then so will $\mathcal{L}_K F$ and in particular, the diffeomorphism induced by $K$ will map solutions of the equations into solutions of the equations. An important example occurs when $C=\star_g$ and $K$ is a Killing vector of $g$.

Suppose that we have a group which acts simply transitively on $M$ by left
actions and which preserves the material configuration, as is the case
for the $\tilde{E}(2)\times\mathbb{R}_t$ symmetry of helical ground
state of the nematic liquid crystal. Then it must be that $C$ may be
written in terms of the left invariant one-forms and their duals as:
\begin{equation}
C = \frac{1}{4}C_{ab}{}^{cd} (\lambda^a \wedge \lambda^b) \otimes (L_c \wedge L_d)
\end{equation}
where $C_{ab}{}^{cd}$ are some \emph{constant} coefficients. Here, indices run over $0,
\ldots,  3$. We can make use of this to write down Maxwell's equations
for a nematic liquid crystal in its helical state. We take
\begin{equation}
F = E_i  \lambda^i \wedge dt + \frac{1}{2} \epsilon_{ijk}B_i \lambda^j \wedge \lambda^k.
\end{equation}
This choice of basis is very similar to the rotating basis chosen by
Peterson, who investigated the electromagnetic field propagating
through a nematic liquid crystal in its ground state
\cite{peterson}. In our case, this choice of basis arises naturally
from the group structure of underlying symmetries. We assume further
that
\begin{eqnarray}
C(\lambda^i \wedge dt) = -\frac{1}{2}\varepsilon_{ij}\epsilon_{jkl} \lambda^k \wedge \lambda^l \nonumber \\
C(\lambda^i \wedge \lambda^j) = \epsilon_{ijk}(\mu^{-1})_{kl} \lambda^l \wedge dt.
\end{eqnarray}
Where $\varepsilon$, $\mu$ have the uniaxial form we previously assumed (\ref{biax}). Maxwell's equations for the electric and magnetic fields take the form
\begin{eqnarray}
L_i(B_i) &=& 0, \qquad \epsilon_{ijk}L_j(E_k) + \frac{\partial
  B_i}{\partial t} -P_{ij}E_j= 0, \nonumber \\
\varepsilon_{ij} L_i(E_j) &=& \rho, \qquad (\mu^{-1})_{kl}\epsilon_{ijk}L_j(B_l) - \varepsilon_{ij}\frac{\partial E_j}{\partial t} -P_{ij}(\mu^{-1})_{jk}B_k= J_i.
\end{eqnarray}
 The matrix $P_{ij}$ has
non-zero components
\ben
P_{11} = P_{22} = 1.
\een
These equations can be separated with the ansatz
\begin{equation}
E_i = e^{i(k_x x + k_y y - \omega t)} f_i(z), \qquad B_i = e^{i(k_x x + k_y y - \omega t)} g_i(z).
\end{equation}
The components $f_3(z), g_3(z)$ are given by a linear combination of
other components, so that the Maxwell equations reduce to a system of
differential equations of the form:
\ben
F'(z) + (\alpha + \beta_1 e^{2 i p z}+ \beta_2 e^{-2 i p z}) F(z) =
0. \label{genmat}
\een
Here $F(z) = (f_1(z), f_2(z), g_1(z), g_2(z))^t$ is a $4$-vector and
$\alpha, \beta_i$ are $4 \times 4$ matrices, given by:
\begin{equation}
\alpha = \left( \begin{array}{cccc}
0 & 1 & 0 & \frac{-i|\kappa|^2}{2 \varepsilon_\perp \mu_{||} \omega} +
i \omega\\
-1 & 0 & \frac{i|\kappa|^2}{2 \varepsilon_\perp \mu_\perp \omega}-i\omega & 0 \\
0  & \frac{i|\kappa|^2}{2\omega} -i\varepsilon_{||}\mu_\perp \omega
& 0 & \frac{\mu_\perp}{\mu_{||}} \\
-\frac{i\mu_{||}|\kappa|^2}{{2\mu_\perp \omega}}+i\varepsilon_{||}\mu_{||} \omega & 0 & -\frac{\mu_{||}}{\mu_\perp} & 0
\end{array} \right )
\end{equation}
and
\begin{equation}
\beta_1 = -\overline{\beta_2} = \frac{\overline{\kappa}^2}{4 \omega}
\left( \begin{array}{cccc} 0 & 0 & \frac{-1}{\varepsilon_\perp
      \mu_\perp} & \frac{-i}{\varepsilon_\perp \mu_{||}} \\ 0 & 0 &
    \frac{-i}{\varepsilon_\perp \mu_\perp}& \frac{1}{\varepsilon_\perp
      \mu_{||}} \\ 1 & i & 0 & 0 \\ i \frac{\mu_{||}}{\mu_\perp} & -
        \frac{\mu_{||}}{\mu_\perp} & 0 & 0
  \end{array} \right )
\end{equation}
where we have introduced $\kappa = k_x + i k_y$. We see that the
Euclidean symmetry of the original problem is still manifest since a
rotation in the $x$--$y$ plane sends $\kappa \to e^{i \theta} \kappa$,
which is cancelled by a suitable shift in the $z$ coordinate. We may
view (\ref{genmat}) as a generalised Mathieu equation. Mathieu's
equation itself may be written in this form with $2 \times 2$
matrices. By Floquet's theorem, the general solution of (\ref{genmat})
will take the form:
\ben
F(z) = e^{i \mu_1 z} h_1(z) + e^{i \mu_2 z} h_2(z) +e^{i \mu_3 z}
h_3(z) + e^{i \mu_4 z} h_4(z)
\een
where $h_i(z) = h_i(z+\pi/p)$ are $4$-vectors. Making use of discrete
symmetries of the equations, one may show that if $\mu$ is a Floquet
exponent, then so is $-\mu$ and $\overline{\mu}$, implying relations
amongst the $\mu_i$. This equation may be
studied using the infinite determinant techniques of Hill, an approach
similar to that of \cite{Dreher}, but that
takes us beyond the scope of the current paper. We hope to address
this issue in future work. Since we have retained the independence of
the magnetic susceptibility and the permittivity, this analysis
applies equally well to magnetic materials with helical phases
\cite{Bak}.

\section{Conclusion}

We have shown that certain properties of the chiral phase of a nematics
liquid crystal are intimately tied to the symmetries it possesses. We
have shown that the Joets-Ribotta metric, which describes the
propagation of extraordinary rays, is a
left-invariant metric on $\tilde{E}(2)$ and we have shown how the
underlying symmetry group can be practically used to understand
properties of waves in such a medium.

We have separated the Hamilton-Jacobi equation and
the wave equation for this metric. The wave equation can be reduced to
Mathieu's equation and the Hamilton-Jacobi equation to the quadrantal
pendulum equation. We have also seen how Maxwell's equations for a
general uniaxial material whose director field lies in a helical
configuration can be reduced to coupled ordinary differential equations generalising Mathieu's
equation via a novel application of the theory of Lie groups. This new
formalism is applicable to the macroscopic Maxwell equations whenever
the medium has a continuous symmetry group. The approach taken
generalises transformation optics to permit non-impedance matched media.

As we have seen even in this simple example, the extraordinary light
rays propagating through a liquid crystal explore a much richer
geometry than the usual flat geometry of light rays in the
vacuum. This opens up the possibility of constructing analogues for
the propagation of light in a gravitational field. In this case the
light rays in the liquid crystal may be mapped onto light rays
propagating in a Bianchi $VII_0$ cosmology \cite{Pontzen} whose spatial sections have
a fixed geometry, but one may imagine more ambitious possibilities.

\appendix

\section{Generalization to Bianchi type $VII_h$}

It is interesting to ask whether the set up above generalises to the
Bianchi type $VII_h$ group. For this section we  set $p=1$,
in order not to clutter up the formulae.

We now define left-invariant one-forms and dual vector fields by 
\bea
\lambda ^3&=&  dz\,,\qquad \qquad \qquad \qquad \qquad  \qquad ~L_3=  \frac{\p}{\p z} \,,\\ 
\lambda ^1 &=& e^{hz} ( \cos z dx + \sin z dy) \,,\qquad L_1= e^{-hz} 
(\cos z \frac{\p}{\p x}  + \sin z  \frac{\p}{\p y}) \,,  \\
\lambda ^2 &=& e^{hz} ( \cos z  dy   - \sin z dx ) \,,\qquad  
L_2 = e^{-hz} ( \cos z \frac{\p}{\p y}-  \sin z)  
\frac{\p}{\p x}  \,.
\eea
The right-invariant one-forms and vectors  fields are
\bea
\rho^3&=&  dz \,, \qquad \qquad \qquad   \qquad R_3 =  
\frac{\p}{\p z} + x \frac{\p}{\p y} -y  \frac{\p}{\p x} -
 h (x \frac{\p}{\p y} +y  \frac{\p}{\p x}    ) \,,\\  
\rho^1&=& dx + (1+h)y dz \,, \qquad R_1= \frac{\p}{\p x} \,,\\
\rho^2&=& dy - (1-h)x dz \,, \qquad R_2= \frac{\p}{\p y} \,.
\eea
The metric \footnote{In what follows $\nee $ and $\no$ will be taken to be 
constant, that is position independent.} 
\ben
\nee ^2 \Bigl ( \lambda _1^2 + \lambda _2^2 + \lambda_3  ^2  \Bigr )   =
n^2_e \Bigl (  dz ^2 + e^{2hz}
( dx ^2 + dy ^2 ) \Bigr )    
\een
is in fact that of hyperbolic three space.
in the upper half space or Poincar\'e patch space model.
Setting 
\ben
e^{hz} = \frac{1}{Z} \,, \quad x= \frac{X}{h} \,, \quad y= \frac{Y}{h}
\een
it becomes
\ben
\frac{\nee ^2}{h^2 Z^2} \Bigl (  dZ^2 +  dX^2 + dY^2 \Bigr )\,,  
\een
and we see that optically we can think of a vertically stratified  
isotropic  medium with Cartesian coordinates $(X,Y,Z)$  and  refractive index
\ben
\frac{\nee }{hZ} \,.
\een
Rays are now circles orthogonal to the plane $Z=0$.

The metric 
\ben
ds ^2_o= \nee ^2 \Bigl ( \lambda _1^2 + \lambda _2^2 + \lambda_3  ^2  \Bigr ) 
+ (\no^2-\nee ^2) \lambda_1^2 
\een
may  thought of as describing a vertical stratified anisotropic medium
with   extraordinary and ordinary refractive indices varying with height
$Z$ in the same way, i.e.  as     
\ben
\frac{\nee }{hZ} \,, 
\qquad {\rm and} \qquad \frac{\no}{hZ} \,\qquad{\rm respectively}\,. 
\een
Such a variation might be due to temperature variation within the material, for example. As before, the wave equation separates but $F(z)$ now satisfies
\ben
\frac{d^2 F}{dz ^2} + 2h \frac{dF}{dz} + 
\Bigl( \omega ^2 \nee ^2 
- \frac{e^{-2hz}}{2}  (1+ \frac{\nee ^2}{ \no^2} )( k_x^2 + k_y^2 )  + 
\frac{e^{-2hz}}{2}  (k_x^2 + k_y^2 )
(1- \frac{\nee ^2}{ \no^2} ) \cos(2z-\theta) 
 \Bigr ) F =0\,.
\een


\begin{thebibliography}{99}

\bibitem{Nakahara}
  M.~Nakahara,
  ``Geometry, topology and physics,''
  Boca Raton, USA: Taylor \& Francis (2003) 573 p.

\bibitem{Frankel}
  T.~Frankel,
  ``The geometry of physics: An introduction,''
  Cambridge, UK: Univ. Pr. (1997) 654 p.

\bibitem{Joets} A.~Joets and R.~Ribotta, A geometrical model for the 
propagation of light rays in an anisotropic inhomogeneous medium
{\it Optics Communications} {\bf 107} (1994) 200-204

\bibitem{Joets1} A.~Joets and R.~Ribotta
Caustics and symmetries in optical imaging. The example of convective flow visualization {\it J. Phys. I France} {\bf 4}  (1994) 1013-1026 

\bibitem{Joets2}A.~Joets and R.~Ribotta, 
 Structure of Caustics Studied Using the Global Theory of Singularities
{\it Europhysics Letters} {\bf 29} (1995) 593-598 

\bibitem{Joets3}A.~Joets and R.~Ribotta, 
Experimental Determination of a Topological Invariant in a Pattern of 
Optical Singularities {\it Phys. Rev. Lett.} {\bf  77} (1996), 1755–1758 

\bibitem{Joets4}A.~Joets and R.~Ribotta,
Defects and interactions with the structures in ehd convection in nematic liquid crystals 
{\it Cellular Structures in Instabilities
Lecture Notes in Physics} {\bf 210}  (1984)  249-262,

\bibitem{peterson}
 M.~A.~Peterson,  Light propagation and light scattering in cholesteric liquid crystals {\it Phys. Rev. A.} {\bf 27} (1983) 520-529

\bibitem{Dreher} R.~Dreher and F.~Meier, Optical properties of
  Cholesteric Liquid Crystals, {\it Phys. Rev. A.} {\bf 8} (1973) 1616-1623






\bibitem{Sethna} J.~P.~Sethna, D.~C.~Wright and N.~D.~Mermin
, Relieving Cholesterol Frustration:The Blue Phase in a Curved Space
{\it Phys Rev Lett} {\bf 51}(1983) 467-470 

\bibitem{Sethna1} J.~ P.~ Sethna, Frustration, and  Curvature, 
Glasses and Cholesteric Blue Phase
{\it Phys Rev Lett } {\bf 51}(1983) 2198 

\bibitem{Sethna2} J.~ P.~ Sethna, Frustration, curvature, and
defect lines in metallic glasses and the cholesteric phase 
{\it Phys Rev} {\bf B \, 31}(1985) 6275 

\bibitem{Bragg} W.~ Bragg, Liquid Crystals, {\it Nature} {\bf 133} 
(1934) 445-456 
 
\bibitem{Collings} P.~J.~ Collings {\it Liquid Crystals} 
Princeton University Press (1990) 

\bibitem{Wright} D.~C. Wright and N. D. Mermin, 
Crystalline liquids:the blue phase {\it Rev Mod Phys} {\bf 61} (1989) 385-432

 


\bibitem{Mermin} N.~D.~ Mermin, 
The topological theory of defects in ordered media, 
{\it Rev Mod Phys} {\bf 51} (1979) 591-648


\bibitem{Kleman} M.~ Kl\'eman, Defects in liquid crystals {\it Rep Prog Phys}
{\bf 52} (1989) 555-654 

\bibitem{Chandra} S~. Chandrasekhar and G.~S.~ Raganath,
The structure and energetics of defects in liquid crystals,
{\it Advances  in Physics}
{\bf 35} (1986) 507--596 













\bibitem{Satiro} C. S\'atiro and F. Moraes, On the deflection of light by 
topological; defects in nematic liquid crystals  {\it Eur. Phys J.} {\bf E \, 25} (2008) 425-429

\bibitem{Pereira}
  E.~R.~Pereira and F.~Moraes,
Flowing Liquid Crystal Simulating the Schwarzschild Metric,
{\it Central European Journal of Physics} to appear
  arXiv:0910.1314 [gr-qc].

\bibitem{Pereira2}
  E.~Pereira and F.~Moraes,
  Diffraction of light by topological defects in liquid crystals,
{\it Liquid Crystals} to appear
  arXiv:0905.1531 [cond-mat.soft].


\bibitem{Bouligand} Y.~Bouligand, B.~ Derrida, V.~ Poenaru, Y.~Pomeau and G.~Toulouse, Distortions with Double Topological Character: The Case of Cholesterics,
{ \it J. de Physique} {\bf 39} (1978) 863-867 
 
\bibitem{Pansu1} B.~Pansu,  E.~Dubois-Violette and R.~Dandoloff, Disclination in the  $S^3$ blue phase
{ \it J. de Physique} {\bf 48} (1987) 305-317
 
\bibitem{Pansu2} B.~Pansu and E.~Dubois-Violette, Textures of $S^3$ blue phase
{ \it J. de Physique} {\bf 48} (1987) 1861-1869

\bibitem{Dandoloff} R.~ Dandoloff and R.~ Mosseri, The Blue Phase: 
from $S^3$ to a Double-Twisted Tube in ${\Bbb R} ^3$, 
{\it Europhysics Letters} {\bf 3}(1987) 1193-1200

\bibitem{Lamb} H.~ Lamb {\it Hydrodynamics} 6th Edition,
 Cambridge University Press
 (1932) p. 176

\bibitem{Birkhoff} Garrett Birkhoff, 
{\it Hydrodynamics, a study in logic, fact, and similitude} 
Princeton University Press, (1950).

\bibitem{Olshanetsky}
  M.~A.~Olshanetsky and A.~M.~Perelomov,
 Classical integrable finite dimensional systems related to Lie algebras,
  {\it Phys. Rept.}   {\bf 71} (1981) 313.

\bibitem{Ramsey} A.~S.~ Ramsey, 
{\it A Treatise on Hydromechanics: Part II Hydrodynamics} 
Fourth Edition G. Bell (1935) p.198 

\bibitem{WW} A Course of Modern Analysis, 4$^{th}$ edition, E.~T.~Whittaker and
  G.~N.~Watson, CUP (1927)

\bibitem{Tamm} J.~E.~ Tamm  {\it J. Russ. Phys-Chemi. Soc} {\bf 56} 2-3 (1924)
284

\bibitem{Plebanski} J.~Plebanski, Electromagnetic Waves in
  Gravitational Fields  {\it Phys. Rev.} {\bf 118}(1960) 1396-1408 


\bibitem{Leonhardt} Ulf Leonhardt, Thomas G. Philbin
Transformation Optics and the Geometry of Light
{\it Prog. Opt.} {\bf 53} (2009)  69-152 
{\tt arXiv:0805.4778v2 } [physics.optics]


\bibitem{Bak} P.~Bak and M.~ H\o gh Jensen, Theory of helical magnetic structure and phase transitions in MnSi and FeGe, {\it J. Phys. } {\bf C\, 13} 
(1980) L881-L885 

\bibitem{Pontzen}
  A.~Pontzen and A.~Challinor,
 Linearization of homogeneous, nearly-isotropic cosmological models,
  arXiv:1009.3935 [gr-qc].

















\end{thebibliography}
\end{document}